# POLAR CONCENTRATION OF ELEMENTS IN TREE LEAVES

James D. Brownridge
Department of Physics, Applied Physics, and Astronomy
Binghamton University, P. 0. Box 6000, Binghamton, New York 13902-6000

**Abstract**

A long-term study of the elements Mg, Al, Si, P, Ca, S, Cl, Fe and Mn in leaves is in progress. The objective of this study is to develop a week-by-week profile of these elements in leaves during several growing seasons. The profile includes the following information: (1) Which elements each tree collects in its leaves. (2) The location in the leaf with the highest concentration, top side, under side or interior. (3) The week during the growing season when each element first appears in the leaves of each tree. (4) The change in the relative concentration from week to week. (5) The source of the element i.e., deposition from the atmosphere or the root system of the tree. This information is profile for each year and will be correlated with environmental conditions for that year. Leaves are collected weekly from first unfolding in early spring until leaf drop in the fall. They are from the 31 trees and 26 species in Broome County, NY. From time to time leaves from most of the 26 species are being randomly collected from trees growing throughout the northeastern US.

**Discussion**

The pignut hickory *(Carya glabra)* is perhaps one of the more spectacular examples of polar concentration and timing of nutrient element uptake in trees. X-ray spectra data is presented in Fig. 1 showing the element content in the surface of leaves collected from unfolding in the spring through leaf dropping in the fall. Leaves at less than three days old were first collected on May 20, 1996. There is little distinction in element content between the topside and underside of the leaves at this early stage of development. However, as the leaves grow, a clear distinction develops between the topside and underside. In pignut hickory the accumulation of Mn begin between May 20 and May 31, Ca between May 31 and June 13, and Al between June 24 and July 1. The pignut hickory is not unique in showing this kind of uptake pattern; however, it does appear to be unique with respect to Al and Mn uptake and location in the leaf. When the leaves of the trees in this study (26 species) first unfold, they look very similar to the pignut hickory. Within days to weeks, however, they begin to "take up" elements in a pattern that appears to be unique to the species and sometimes unique to a specific tree within a species even when branches and roots overlap.

Leaves are collected from the same trees at regular intervals (7-10 days) from the date of first appearance until leaf dropping in the fall. Usually five leaves are collected and immediately vacuum dried. A small sample of each leaf (about 2 cm$^2$) is placed in a specially designed[1, 2] X-ray fluorescence spectrometer, and X-ray spectra of elements in the leaf surface are produced. The leaf sample is then rotated 180° and the procedure is repeated. This is key to the determination of element content in the top and undersides of leaves to a depth of less than 2µm. Pulverizing a sample of a leaf and analyzing the powder, the usual method[3, 4], is used to obtain information about the interior of leaves. Figure 2 is a typical example of the type of information that is both lost and gained when a leaf is pulverized. It is clear that the Ca/Al and the Ca/Mn ratios are dramatically different for each analysis. All three analyses were done on the same leaf sample and normalized to the same count in a region of the spectra around 3 keV. When this is done, the relative concentration of each element in the leaf may be inferred from the number of counts in each peak. It is clear that most of the Al is in the topside layer and virtually all of the Mn is in the underside layer of the leaf. Potassium, Sulfur, and Chlorine are mostly concentrated inside the leaf[5].

**References**

1. Brownridge, J. D. Pyroelectric X-ray Generator. Nature 358, 287-289 (1992).

2. Brownridge, J. D. and Raboy, S. Investigation of Pyroelectric Generation of X Rays. Submitted and in revision (Sep. 1998) J. Appl. Phys.

3. Kocman, V., Peel, T. E. and Tomlinson, G. H. Rapid Analysis of Micro Nutrients in Leaves and Vegetation by Automatic X-ray Fluorescence Spectrometry (A Case Study of an Acid-Rain Affected Forest). Commun. Soil Sci. Plant Anal. 22, 2063-2075 (1991).

4. Johoansson, M. B. The Chemical Composition of Needle and Leaf Litter from Scots Pine, Norway Spruce and White Birch in Scandinavian Forest. Forestry **68, 49-62** (1995).

5.  For periodic updates see:  http://www.binghamton.edu/physics/brownridge.html

jdbjdb@binghamton.edu

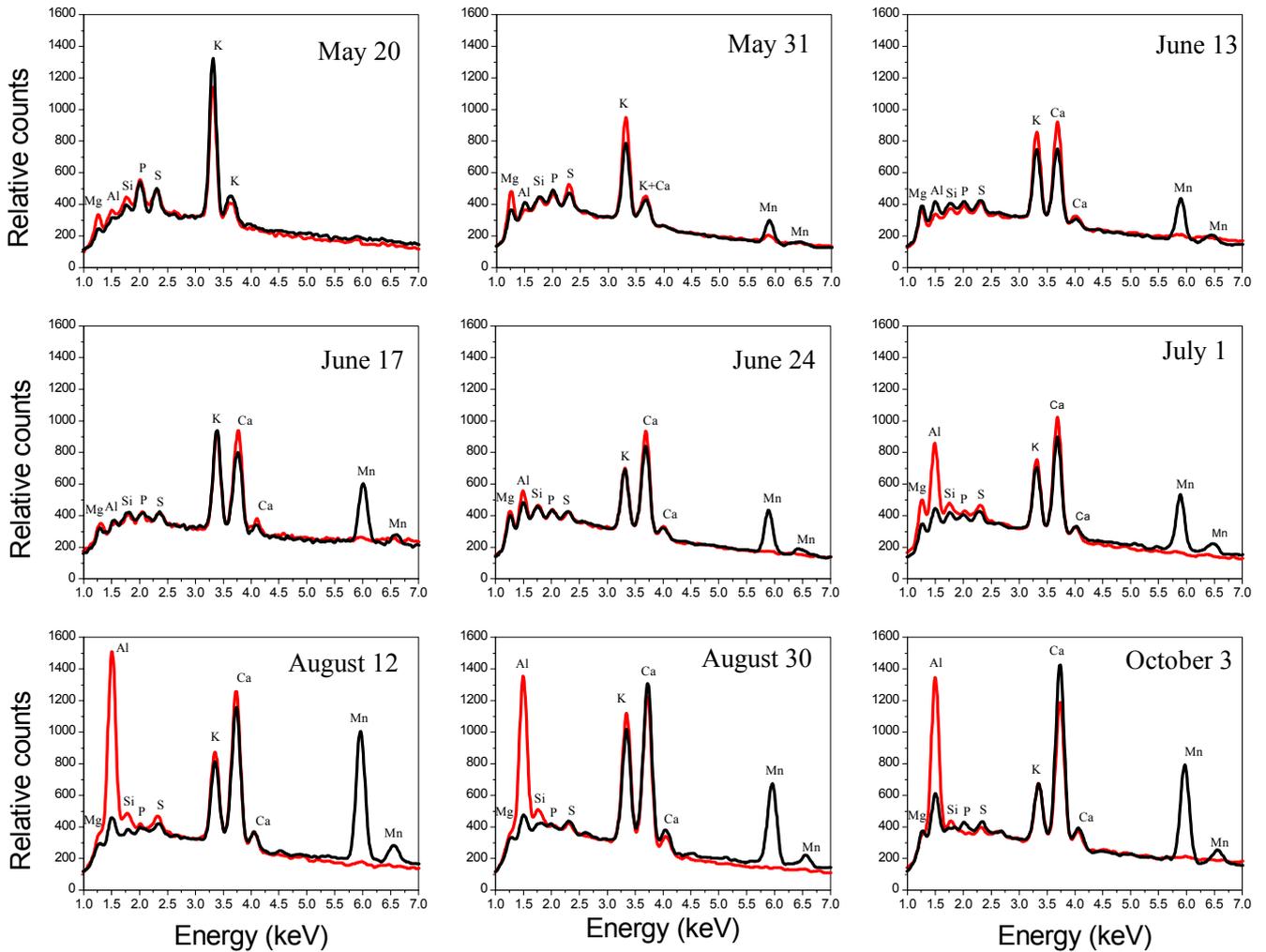

Fig. 1. X-ray spectra of the elements on the topside (red) and underside (black) of leaves collected from a pignut hickory. Elements with an atomic number below 12 cannot be detected with the equipment used in this study. Leaves were collected on the date shown in each sub-figure. The spectra in each subfigure show the approximate relative concentration of each element in the surface layer of leaves on the tree at the time of collection. When the name of an element appears twice it is a denotation of the Kα and K$_β$ X-rays for that element, the Kα is the larger peak. Note that the K$_β$ of potassium and the Kα of calcium overlap. The equipment used in this study does not resolve the Kα, K$_β$ X-rays of elements below potassium.

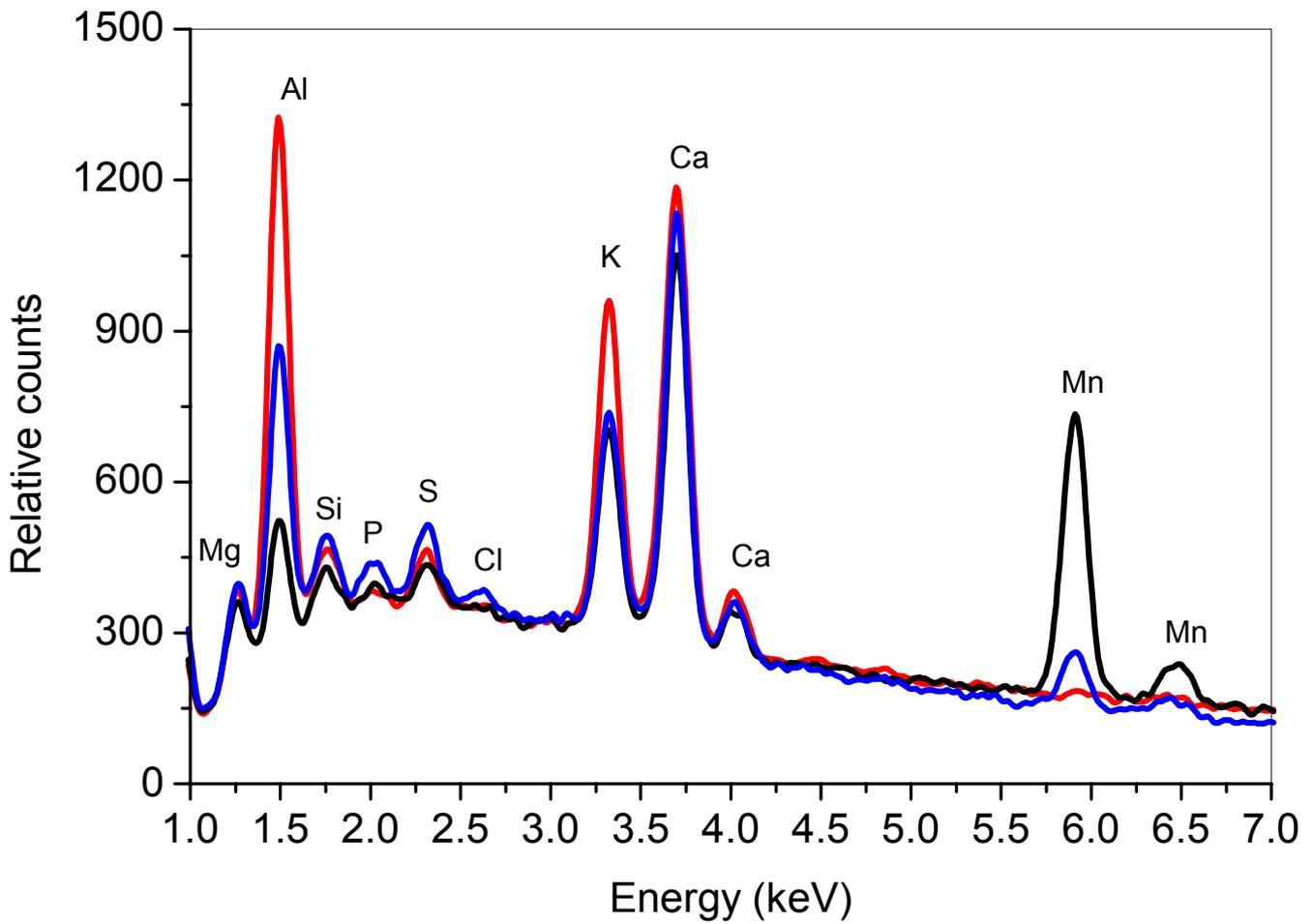

Fig. 2. X-ray spectra of the topside (**red**), underside (**black**), and powdered leaf (**blue**) collected on August 12. Concentrations below about 1μg /cm$^2$ and elements with atomic numbers below 11 are not detectable.

I